# Anomalous spin-orbit torque switching due to field-like torque-assisted domain wall reflection

Jungbum Yoon,[1*†] Seo-Won Lee,[2*] Jae Hyun Kwon,[1] Jong Min Lee,[1] Jaesung Son,[1] Xuepeng Qiu,[4] Kyung-Jin Lee,[2,3‡] and Hyunsoo Yang[1‡]

[1]*Department of Electrical and Computer Engineering, National University of Singapore, 117576, Singapore*

[2]*Department of Materials Science and Engineering, Korea University, Seoul 02841, Republic of Korea*

[3]*KU-KIST Graduate School of Converging Science and Technology, Korea University, Seoul 02841, Republic of Korea*

[4]*Shanghai Key Laboratory of Special Artificial Microstructure Materials & School of Physics Science and Engineering, Tongji University, Shanghai 200092, China*

*These authors contributed equally to this work.

†Present address: Center for Nanometrology, Korea Research Institute of Standards and Science, Daejeon 34113, Republic of Korea

‡Corresponding author. E-mail: eleyang@nus.edu.sg (H.Y.); kj_lee@korea.ac.kr (K-J. L.).

**Spin-orbit torques (SOT) allow the electrical control of magnetic states. Current-induced SOT switching of the perpendicular magnetization is of particular technological importance. The SOT consists of damping-like and field-like torques so that the efficient SOT switching requires to understand combined effects of the two torque-components. Previous quasi-static measurements have reported an increased switching probability with the width of current pulses, as predicted with considering the damping-like torque only. Here we report a decreased switching probability at longer pulse-widths, based on time-resolved measurements. Micromagnetic analysis reveals that this anomalous SOT switching results**

**from domain wall reflections at sample edges. The domain wall reflection is found to strongly depend on the field-like torque and its relative sign to the damping-like torque. Our result demonstrates a key role of the field-like torque in the deterministic SOT switching and notifies the importance of sign correlation of the two torque-components, which may shed light on the SOT switching mechanism.**

**INTRODUCTION**

Spin-orbit coupling is able to convert charge currents to spin currents (*1-8*). The ability to generate spin currents without the help of ferromagnets constitutes a core building block of an emerging research field, *spin-orbitronics*, which pursues the use of spin-orbit coupling as a spin-current source in spintronic devices. When the spin current due to spin-orbit coupling is absorbed by a ferromagnet, it exerts a spin-orbit torque (SOT) on the ferromagnet. The SOT is able to switch magnetization (*9, 10*) and to induce fast domain wall motion (*11-13*) in ferromagnet/heavy metal bilayers. The SOT switching of the perpendicular magnetization is of particular technological relevance as perpendicular magnetic random access memories have a better scalability than in-plane ones.

From the viewpoint of fundamental physics as well as applications, it is of critical importance to understand the detailed SOT characteristics and consequent magnetization dynamics. The microscopic origin of SOT remains under debate (*9, 10, 14-26*), but it is commonly decomposed into two mutually orthogonal vector components, the damping-like torque (DLT) and field-like torque (FLT). SOT-induced magnetization dynamics is described by the Landau-Lifshitz-Gilbert equation including the DLT and FLT terms:

$$\frac{d\hat{\mathbf{m}}}{dt} = -\gamma\mu_0 \hat{\mathbf{m}} \times \mathbf{H}_{\text{eff}} + \alpha \hat{\mathbf{m}} \times \frac{d\hat{\mathbf{m}}}{dt} + \gamma\tau_{\mathbf{d}} \hat{\mathbf{m}} \times (\hat{\mathbf{m}} \times \hat{\mathbf{y}}) + \gamma\tau_{\mathbf{f}} \hat{\mathbf{m}} \times \hat{\mathbf{y}}, \qquad (1)$$

where $\gamma$ is the gyromagnetic ratio, $\hat{\mathbf{m}}$ is the unit vector along the magnetization, $\mu_0\mathbf{H}_{\text{eff}}$ is the effective magnetic field including the external, anisotropy, magnetostatic, and exchange fields, and $\alpha$ is the damping constant. $\tau_d$ ($=(\hbar/2e)(J/M_sd)c^{\parallel}$) and $\tau_f$ ($=(\hbar/2e)(J/M_sd)c^{\perp}$) respectively describe the magnitudes of DLT and FLT in the unit of magnetic field, where $J$ is the current density, $M_s$ is the saturation magnetization, $d$ is the thickness of ferromagnet, $c^{\parallel}$ and $c^{\perp}$ are respectively the DLT and FLT efficiencies, and $\hat{\mathbf{y}}$ is the unit vector perpendicular to both the current direction and the inversion asymmetry direction (i.e., thickness direction; see coordinate system in Fig. 1A). From Eq. (1), one finds that the two torque-components affect magnetization dynamics in a distinctly different way: the DLT directs the magnetization towards the *y*-axis, while the FLT induces magnetization precessions around the *y*-axis.

Most previous SOT-switching studies have considered the DLT as a main driving source but ignored the FLT. When considering the DLT only, the switching trajectory in the macrospin approximation is expected to be simple without magnetization precessions (*27*). This DLT-dominated switching leads to an increased switching probability with the current pulse-width, consistent with previous quasi-static measurements (*28, 29*) and also in accordance with a common belief. In some ferromagnet/heavy metal bilayers (e.g., Ta-based bilayers), however, the FLT is significant (*30-32*). For a sizable FLT, it has been theoretically predicted that the magnetization precession induced by the FLT complicates magnetization dynamics especially for the case where the sign of FLT is the opposite to that of DLT (in our sign convention; see Eq. (1)) (*33-35*). In this respect, it is important to experimentally investigate the role of FLT in the SOT switching. We note that previous quasi-static measurements based on Hall bar detection (*9, 10, 28, 29*) or magneto optical Kerr effect (MOKE) microscopy (*36*) would be unable to capture the core effect of FLT because the FLT may induce fast dynamics (i.e., magnetization precessions around the *y*-axis). To overcome this limitation, it is essential to perform time-resolved measurements (*37, 38*), which provide an important step towards a better understanding of SOT-induced magnetization dynamics.

Here, we report SOT-induced magnetization dynamics in time domain by time-resolved (TR)-MOKE measurements for Ta/CoFeB/MgO heterostructures with perpendicular magnetic anisotropy. This Ta-based structure has a large FLT (i.e., $|\tau_f/\tau_d| \approx 4$) whose sign is the opposite to that of DLT (*32*), so that it allows a detailed study of FLT effect on the SOT switching. The temporal evolution of the magnetization is detected by the stroboscopic pump-probe technique with an electrical pulse generator (pump) and picosecond laser (probe) (see section S1), as shown schematically in Fig. 1A. We observe an anomalous SOT-induced switching behavior, in which the switching probability increases at short current pulses but decreases at longer pulses. Based on micromagnetic simulations, we interpret this anomalous switching behavior as a consequence of FLT-assisted domain wall reflection at sample edges.

**RESULTS**

**Time-resolved measurements of perpendicular magnetization switching by SOT**

We first perform a static polar MOKE measurement using an in-plane dc current *I* with an external in-plane magnetic field $\mu_0 H_x$ along the *x*-axis to examine the dc SOT switching characteristics of the Ta (6 nm)/CoFeB (0.8 nm)/MgO (2 nm) sample (device 1). The pattern size of device 1 is 3 × 3 $\mu m^2$, which is large enough to detect SOT-induced change in the MOKE signal. Figure 1B shows MOKE signals as a function of dc current with various in-plane magnetic fields. As a polar MOKE signal probes the average *z*-component of CoFeB magnetization (<$M_z$>) in a laser spot, where a high (low) signal corresponds to <$M_z$>-up (-down) magnetic state, the hysteretic curves show SOT-induced deterministic magnetization switching. The switching polarity is determined by the direction of current and $H_x$ (9, 10): the up-to-down switching occurs in a positive (negative) current with a negative (positive) $H_x$. As shown in Fig. 1C, the dc switching current density $J_c$ decreases with increasing $H_x$, in agreement with previous reports by Hall bar experiments (9, 20).

We next carry out time-resolved measurements by injecting a current pulse with various pulse-widths (i.e., pulse-width $t_{pw} \leq 5$ ns, current density $J = 5.2 \times 10^{11}$ A m$^{-2}$, and $\mu_0 H_x = -168$ mT) and detecting the magnetic state in time domain through TR-MOKE signal (Fig. 2A). The current pulse is turned on at $t = 0$ (i.e., current-on) and turned off at a time indicated by a red triangle in each curve (i.e., current-off). The horizontal dashed lines are to guide the maximum change in MOKE signal (~ 7 µV), corresponding to full magnetization switching from the up to down state.

For short pulses ($t_{pw} \leq 1.6$ ns), the MOKE signal change is smaller than 7 µV, indicating that the current pulse is too short to switch the magnetization. For an increased $t_{pw}$ to 1.8 ns, a complete switching is achieved as evidenced by the signal change of ~ 7 µV (by defining the final magnetic state at $t = 8$ ns). When $t_{pw} > 2.5$ ns, however, we clearly observe an anomalous temporal change in the TR-MOKE signal: it decreases in the initial time stage (< 2 ns), but increases back even before the current pulse is turned off.

In order to clarify the anomalous SOT switching behavior, we compare two cases, $t_{pw} =$ 1.8 and 5.0 ns, in Fig. 2B. For $t_{pw} = 1.8$ ns, which corresponds to a normal switching, the TR-MOKE signal decreases monotonously when the current is on, indicating that $<M_z>$ changes monotonously from the up to down state by SOT. This decreased signal is maintained even after the current is off so that the up-to-down switching is completed. For $t_{pw} = 5.0$ ns, which corresponds to an anomalous switching, on the other hand, the signal decreases until $t = 2$ ns and then increases back. We note that this anomalous increase in the signal is present even before the current is off. It implies that the SOT is responsible not only for the initial decrease in the MOKE signal (i.e., switching) but also for the anomalous increase in the signal (i.e., switching-back). This anomalous switching-back phenomena are observed for all cases with $t_{pw} > 2.0$ ns, as summarized in Fig. 2C, which shows the switching probability ($P_{SW}$) as a function of $t_{pw}$, where $P_{SW} = (1 - V_{MOKE}(t = 8$ ns$) / 7$ μV$) \times 100$ (%).

We next show that the anomalous switching-back phenomena become more pronounced as the current amplitude increases. We perform TR-MOKE measurements at various current densities and pulse-widths for a structure of Ta (3 nm)/CoFeB (1.2 nm)/MgO (2 nm) with a pattern size of $3 \times 6$ μm$^2$ (device 2). As an example, Fig. 3A shows TR-MOKE signals normalized by the maximum signal change as a function of the time at various current densities for $t_{pw} = 30$ ns. Focusing on the normalized MOKE signal at $t = 45$ ns (i.e., after the current is off), we find that a higher current density causes a more switching-back. Figure 3B summarizes $P_{SW}$ versus $t_{pw}$ at various current densities. It clearly shows that the switching-back phenomena become more noticeable at higher current densities.

**Anomalous SOT switching due to domain wall reflections at edges: micromagnetic analysis**

In order to understand the anomalous switching-back phenomena, we perform micromagnetic simulations at 0 K (Fig. 4; see Methods). We use parameters of magnetic properties and SOTs (DLT ($c^{\parallel} = -0.07$) and FLT ($c^{\perp} = +0.28$)), deduced from magnetometer (see section S2) and harmonic Hall measurements for Ta/CoFeB/MgO samples (*32*). Figure 4A shows temporal evolutions of normalized $<m_z>$ at various current pulse-widths for $J = 15 \times 10^{11}$ A m$^{-2}$ and $\mu_0 H_x = -200$ mT. The current-off time is depicted as a vertical dashed line for each case. For $t_{pw} = 1.5$ and 1.6 ns, the current pulse is too short to switch the magnetization, whereas for $t_{pw} = 1.7$ ns, a full switching is achieved. Interestingly, the switching-back is observed for a longer pulse ($t_{pw} = 1.8$ ns): $<m_z>$ returns back to the initial state ($<m_z>$ = +1) after the current is turned off. Figure 4B shows a switching parameter $P$ as a function of $t_{pw}$, where $P$ is defined as "1" ("0") for the switching (no-switching) event. We find that the switching-back (equivalently no-switching event) is not unique to the case for $t_{pw}$ = 1.8 ns, but appears in a somewhat oscillatory manner for $t_{pw}$ > 1.8 ns. We note that the oscillation is not periodic. This non-periodic oscillatory switching obtained from the zero-temperature calculation (Fig. 4B) can explain the decreased switching probability at longer pulses observed in the room-temperature measurement (Fig. 2C), as the thermal effect randomizes the oscillatory switching dynamics. In section S3, we show micromagnetic simulation results for detailed oscillatory switching dynamics.

Micromagnetic simulations reveal that the switching-back phenomena originate from the domain wall reflection at sample edges in the presence of current (thus SOT), as discussed below. We show temporal evolutions of $m_z$ for $t_{pw}$ = 1.7 ns and 1.8 ns, corresponding to the switching (Fig. 4D) and switching-back (Fig. 4E), respectively. For both cases, in the initial time stage ($t$ < 1.7 ns), a reversed domain is nucleated at a corner and then expands isotropically. For $t_{pw}$ = 1.7 ns, the current is turned off at the moment when the domain wall arrives at sample edges. In this case, the domain wall keeps moving in the same direction as before due to the inertia (*39-41*), even after the current is turned off. As a result, a full switching is achieved. For $t_{pw}$ = 1.8 ns, on the other hand, the current is still turned on when the domain wall arrives at sample edges. In this case, the domain wall is reflected from the edges and moves in the opposite direction, leading to a switching-back. Therefore, the domain wall reflection at sample edges in the presence of SOT is key to explain the anomalous switching-back phenomena.

The domain wall reflection at sample edges also explains the current-dependent switching-back behavior shown in Fig. 3. Figure 4C shows temporal evolutions of $<m_z>$ at various current densities when $t_{pw}$ = 1.2 ns and $\mu_0 H_x$ = −200 mT. At a low current density, a reversed domain is not nucleated ($J$ = 14 × $10^{11}$ A m$^{-2}$) or the domain wall is unable to reach the sample edges ($J$ = 16 × $10^{11}$ A m$^{-2}$). At an increased current density ($J$ = 18 × $10^{11}$ A m$^{-2}$), a full switching occurs. At a higher current density ($J$ = 20 × $10^{11}$ A m$^{-2}$), however, <$m_z$> initially decreases but returns back to the initial state, because of the domain wall reflection (see section S3). Figure 4F shows the switching parameter $P$ as a function of $t_{pw}$ at various current densities. Similar to the results shown in Fig. 4B, the switching-back appears in a somewhat oscillatory manner. Importantly, the no-switching event ($P$ = 0) becomes more frequent for a higher current density when $t_{pw}$ exceeds a threshold to enable the switching. This behavior is consistent with experimental observations (Fig. 3), in which the switching-back phenomena become more pronounced as the current amplitude increases.

**Effect of FLT on the domain wall reflection: collective coordinate analysis**

In order to understand the effect of FLT on the domain wall reflection, we investigate domain wall dynamics based on a semi-one-dimensional micromagnetic model. We first show the domain wall moving along the bottom edge in a two-dimensional sample (Fig. 4D and E) where a Néel-type domain wall is stabilized by $H_x$ at this edge. Figure 5A shows temporal evolutions of the domain wall position at various ratios of the FLT to DLT (i.e., $\tau_f/\tau_d = c^\perp/c^\parallel$) for $J$ = 6 × $10^{11}$ A m$^{-2}$, and $\mu_0 H_x$ = −200 mT. Here we fix the DLT efficiency $c^\parallel$ as −0.07 and vary the FLT efficiency $c^\perp$. For all cases, the domain wall is reflected at the edge (located at 2 μm), exhibits a backward motion for a while, and then moves back to the edge again. After several reflections, the domain wall eventually annihilates at the edge and the switching is completed. This domain wall reflection is understood by the reflection of a transverse wave at a fixed end. It is well known when a transverse wave is reflected at a fixed end, its phase changes by $\pi$ (Fig. 5B). Because a domain wall can be decomposed into transverse spin waves, the phase change corresponds to a change in the domain wall angle $\phi$ upon reflection (see Fig. 5C for schematics). The most important feature in Fig. 5A is that the distance $\Delta q$ (defined as a positive value) for the backward domain wall motion strongly depends on the magnitude and sign of $\tau_f/\tau_d$. For a positive $\tau_f/\tau_d$, $\Delta q$ is small and the domain wall annihilates soon after it reaches the sample edge. For a negative $\tau_f/\tau_d$, on the other hand, $\Delta q$ is large, which in turn causes a noticeable switching-back behavior as shown in Fig. 4E. For a domain wall moving along the left edge (parallel to the y-axis), where the Bloch-type domain wall is stabilized by $H_x$, it experiences a similar reflection process because of the same symmetry of domain wall angle and SOT (see section S4).

We adopt the collective coordinate approach for the domain wall position $q$ and domain wall angle $\phi$ (*42-46*) (see section S5) to explain the dependence of $\Delta q$ on the magnitude and sign of $\tau_f/\tau_d$. We define three domain wall angles $\phi_{ref}$, $\phi_0$, and $\phi_{std}$ (see Fig. 5C and D): $\phi_{ref}$ is the angle just after the reflection, $\phi_0$ is the angle at which the backward domain wall motion reverses to the forward motion, and $\phi_{std}$ is the angle for the steady-state motion before the reflection. We note that $\phi_{ref} = 2\phi_M - \phi_{std}$, where $\phi_M (\equiv \pi - \tan^{-1}(\tau_f / H_x))$ describes the domain tilting in the film plane (see Fig. 5C). Since the edge acts as a fixed end, the domain wall component transverse to the azimuthal angle $\phi_M$ of the magnetization is reversed upon reflection. In other words, $\phi_{std}$, the angle of incoming domain wall, can be rewritten as $\phi_{std} = \phi_M + (\phi_{std} - \phi_M)$, where the first (second) term is longitudinal (transverse) to the domain angle $\phi_M$. Upon reflection, only a transverse component is reversed (i.e., changes its sign), whereas a longitudinal component is conserved. Therefore, $\phi_{ref} = \phi_M - (\phi_{std} - \phi_M) = 2\phi_M - \phi_{std}$.

From the collective coordinate approach, we obtain for $-H_x > \tau_f > 0$ (see section S5)

$$\phi_{std} = \pi - \tan^{-1}\left(\frac{\tau_d/F_- + \alpha\tau_f}{\alpha H_x}\right), \tag{2}$$

$$\phi_{ref} = \pi - 2\tan^{-1}\left(\frac{\tau_f}{H_x}\right) + \tan^{-1}\left(\frac{\tau_d/F_- + \alpha\tau_f}{\alpha H_x}\right), \tag{3}$$

$$\phi_0 = \pi + \tan^{-1}\left[\frac{F_+ \alpha\tau_d - \tau_f}{H_x}\right], \tag{4}$$

where $F_\pm = 1 \pm 2h\xi/\sqrt{1-h^2}$, $\xi = \tan^{-1}((1-h)/\sqrt{1-h^2})$, $h = \sqrt{H_x^2 + \tau_f^2}/H_{K,\text{eff}}$, and $H_{k,\text{eff}}$ is the effective anisotropy field. We find that Eqs. (2-4) describe general tendencies of the numerically obtained three domain wall angles with respect to $\tau_f/\tau_d$ (Fig. 5E). Some disagreement for $\phi_{ref}$ can be attributed to the dynamically distorted domain wall profile just after the reflection.

From the collective coordinate approach with assuming a small damping, an approximate $\Delta q$ is given as (see section S5 for details),

$$\Delta q \approx \sqrt{1-h^2}\lambda\frac{H_x}{\tau_d}\ln\left(\frac{\cos\phi_0}{\cos\phi_{ref}}\right) + \sqrt{1-h^2}\lambda\frac{\tau_f}{\tau_d}(\phi_{ref} - \phi_0), \tag{5}$$

where $\lambda$ is the domain wall width. Figure 5F shows that Eq. (5) describes the numerical results of the dependence of $\Delta q$ on $\tau_f/\tau_d$ qualitatively. We note that the first term of Eq. (5) dominates over the second term (Fig. 5F). Therefore, the dependence of $\Delta q$ on $\tau_f/\tau_d$ is mostly governed by $\cos\phi_0/\cos\phi_{ref}$. We also note that $\phi_{ref}$ changes more rapidly with $\tau_f/\tau_d$ than $\phi_0$ (Fig. 5E). Therefore, the FLT-dependence of $\phi_{ref}$ is key to understand a large backward domain wall motion for a negative $\tau_f/\tau_d$. From Eq. (3), the FLT-dependence of $\phi_{ref}$ is described by $-2\tan^{-1}(\tau_f/H_x)$ (thus $\phi_M$) in a small damping approximation. It means that the FLT affects the backward domain wall motion through its effect on $\phi_M$.

**DISCUSSION**

We demonstrate that the FLT has a crucial role in the SOT switching and causes the anomalous switching-back phenomena when it is large and its sign is the opposite to that of DLT. Our result suggests that not only DLT but also FLT should be carefully examined to achieve the deterministic SOT switching. Furthermore, our result naturally raises a question on the microscopic origin of SOT: what determines the sign correlation between DLT and FLT? To our knowledge, the sign product of DLT and FLT in all previous experiments has been negative (in our sign convention) except for a data point in ref. 30, of which the spin-orbit effective field is too small to unambiguously determine the sign product. In our previous work (*26*), for instance, the signs of DLT and FLT vary with the oxidation of a ferromagnet, but the sign product is always negative. If this fixed sign product is true indeed, it indicates that a single dominant mechanism is responsible for both DLT and FLT. It is in contrast to the currently widely-accepted argument that the DLT (FLT) originates from the bulk spin Hall effect in a heavy metal layer (the Rashba effect at the ferromagnet/heavy metal interface). It is also worthwhile noting that in theories for the bulk spin Hall mechanism, the sign product is positive (*20*) for a positive imaginary part of the spin-mixing conductance (*47*). On the other hand, in theories for the Rashba mechanism, the sign product is negative (*17, 25*).

We have demonstrated the anomalous switching-back phenomena in rather large samples (a few $\mu m^2$), which is required to get detectable MOKE signals. We note that, however, these results could be applicable to nano-sized samples (a few tens of nm), as long as the sample size is larger than the domain wall width because the domain wall dynamics is key to the anomalous switching-back phenomena.

## MATERIALS AND METHODS

**Sample fabrication and TR-MOKE measurements**

The magnetic films of Ta (6 nm)/Co$_{40}$Fe$_{40}$B$_{20}$ (0.8 nm)/MgO (2 nm)/SiO$_2$ (3 nm) and Ta (3 nm)/Co$_{40}$Fe$_{40}$B$_{20}$ (1.2 nm)/MgO (2 nm)/SiO$_2$ (3 nm) are deposited on the thermally oxidized silicon substrates by magnetron sputtering with a base pressure of $< 2 \times 10^{-9}$ Torr at room temperature, and patterned into a square of $3 \times 3$ μm$^2$ or $3 \times 6$ μm$^2$. For a short pulse excitation, the ground-signal-ground (GSG) coplanar waveguide is patterned by electron-beam lithography and deposited with Ta (3 nm)/Cu (75 nm)/Ta (4.5 nm). In the stroboscopic pump-probe experiments, the pulse generator (pump) and the picosecond laser (probe) controller are synchronized by the pattern generator with a triggering frequency of 100 kHz, and each data point of TR-MOKE signal corresponds to the average of 60000 events. The spot diameter of the laser beam is 2 μm. The reflected laser beam is measured by a balanced photodetector to obtain the MOKE signal. Measurements are carried out on three devices and show similar results.

**Micromagnetic simulations**

Micromagnetic simulations are carried out by numerically solving Eq. (1) at zero temperature. Following parameters are used: $M_s = 1.0 \times 10^6$ A m$^{-1}$, the exchange stiffness constant $A_{ex} = 1.0 \times 10^{-11}$ J m$^{-1}$, perpendicular anisotropy constant $K_\perp = 0.9 \times 10^6$ J m$^{-3}$, $\alpha = 0.02$, and $c^\parallel = -0.07$, and $c^\perp = +0.28$. The sample dimension for Fig. 4 is $200 \times 200 \times 1$ nm$^3$ and the unit cell size is $2 \times 2 \times 1$ nm$^3$. The sample dimension for Fig. 5 is $2000 \times 50 \times 1$ nm$^3$ and the unit cell size is $2 \times 50 \times 1$ nm$^3$. For the current pulse, both rise and fall times are 100 ps. In our sign convention, a negative DLT efficiency ($c^\parallel < 0$) induces an up-to-down switching for $J > 0$ and $H_x < 0$. For two-dimensional micromagnetic simulations (Fig. 4), we introduce an artificial defect at the bottom-left corner and consider local demagnetization fields in order to mimic a domain wall nucleation at room temperature.

**SUPPLEMENTARY MATERIALS**

section S1. Stroboscopic pump-probe MOKE experiments using a picosecond laser.

section S2. Characterization of magnetic films.

section S3. Oscillatory SOT-induced magnetization switching: micromagnetic simulations.

section S4. Domain wall moving along the left edge.

section S5. Backward motion of a domain wall reflected at an edge.

fig. S1. Stroboscopic pump-probe magneto-optical Kerr effect (MOKE) set-up.

fig. S2. Vibrating sample magnetometry (VSM) and magneto-optical Kerr effect (MOKE) measurements.

fig. S3. Time-varying *z*-component of the magnetization and its configurations.

fig. S4. Domain wall types formed in the two-dimensional sample.

fig. S5. Temporal evolutions of domain wall position *q* for the Bloch type-wall.

References (*42-50*)

**Funding:** This work was supported by the National Research Foundation (NRF), Prime Minister's Office, Singapore, under its Competitive Research Programme (CRP award no. NRFCRP12-2013-01). K.J.L. was supported by the National Research Foundation of Korea (NRF) grant funded by the Korea government (MSIP) (2011-0027905, 2015M3D1A1070465).

**Author contributions:**

J.Y., J.H.K., and H.Y. initiated this work. J.Y. and X.Q. deposited films. J.Y., J.H.K., and J.S. fabricated devices. J.Y. carried out switching measurements. J.M.L. characterized films. S.-W.L., J.Y., and K.-J.L. performed theoretical and numerical studies. All authors discussed the results. J.Y, S.-W.L., K.-J.L., and H.Y. wrote the manuscript. H.Y. supervised and led the project.

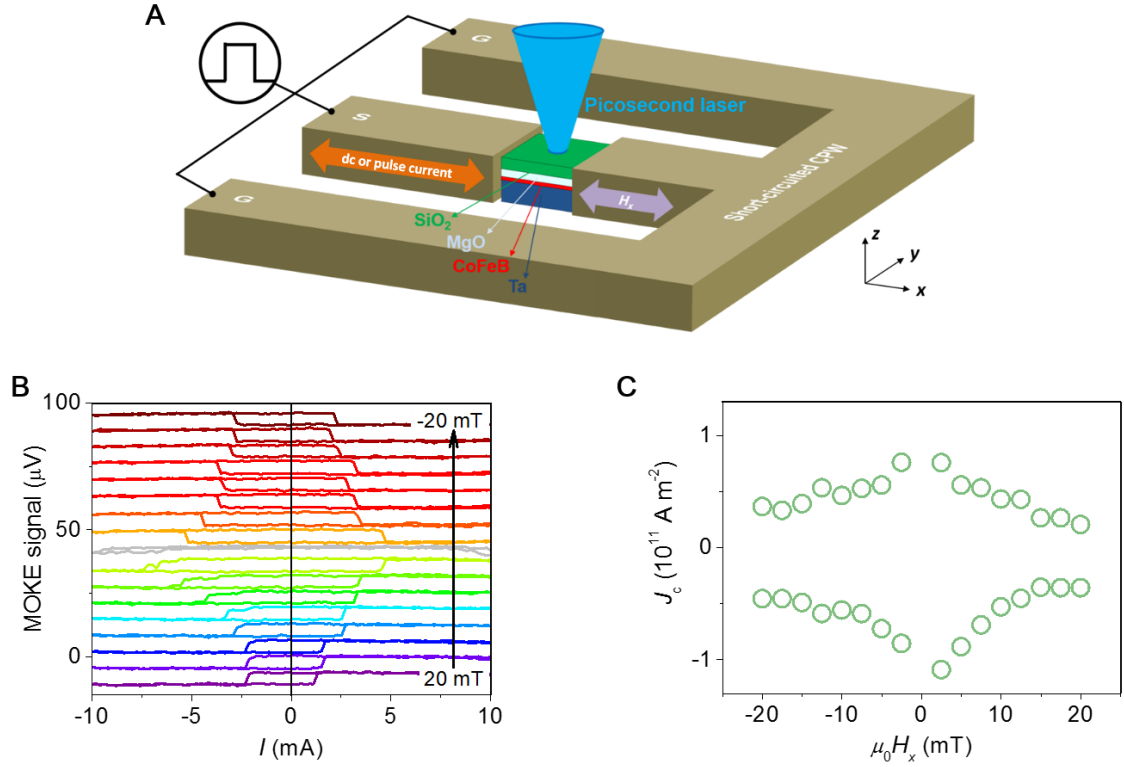

**Fig. 1. TR-MOKE experimental set-up and dc current induced magnetization switching in Ta/CoFeB/MgO.** (**A**) Schematic illustration of TR-MOKE measurements. $\mu_0 H_x$ is the external magnetic field. The dc or pulse current is applied along the *x*-axis. The picosecond laser is shined as a probe beam. The patterned perpendicular anisotropy Ta (6 nm)/CoFeB (0.8 nm)/MgO (2 nm) square is connected to a ground (G)-signal (S)-ground (G) coplanar waveguide. (**B**) dc current induced magnetization switching with various $\mu_0 H_x$. The data are shifted vertically for clarity. (**C**) dc switching current density $J_c$ versus $\mu_0 H_x$.

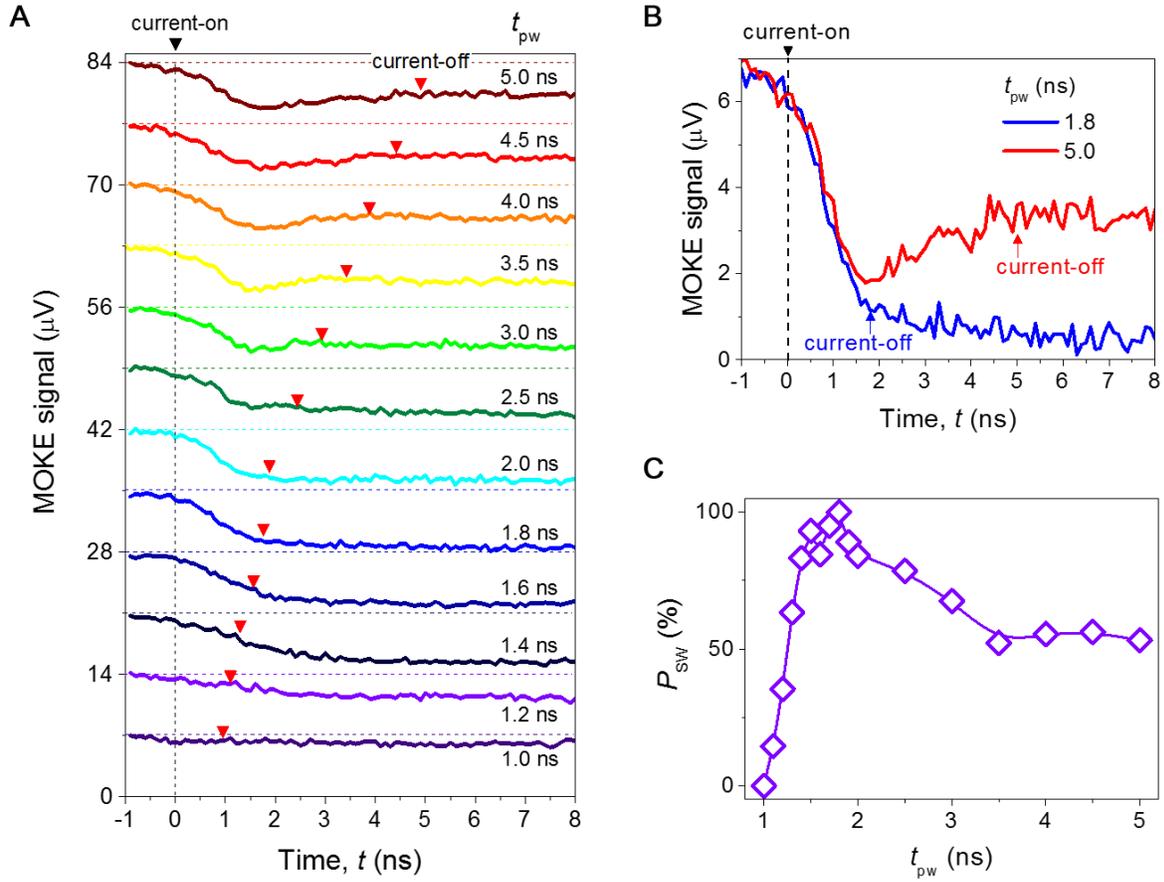

**Fig. 2. TR-MOKE measurements of SOT-induced perpendicular magnetization switching in Ta/CoFeB/MgO.** (**A**) Temporal evolutions of TR-MOKE signals corresponding to the average $z$-component of magnetization ($M_z$) in an applied current density ($J = 5.2 \times 10^{11}$ A m$^{-2}$) of various pulse widths ($t_{pw}$) from 1 to 5 ns, for $\mu_0 H_x = -168$ mT. The data are shifted vertically for clarity. The current pulse starts at $t = 0$ and the end of the current pulse is indicated as a red triangle in each curve. The horizontal dashed lines are to guide the maximum change in MOKE signal (~7 $\mu$V), corresponding to the full magnetization switching from the up to down state. (**B**) Time-varying MOKE signal for $t_{pw} = 1.8$ and 5.0 ns. (**C**) Switching probability ($P_{sw}$) as a function of $t_{pw}$, extracted from Fig. 2A.

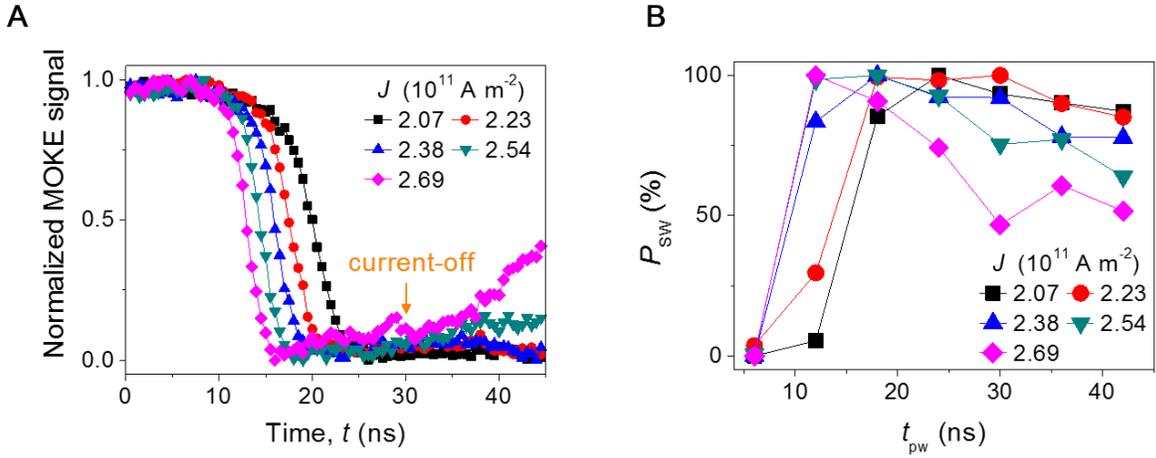

**Fig. 3. Current-dependence of SOT-induced magnetization switching.** The sample structure is Ta (3 nm)/CoFeB (1.2 nm)/MgO (2 nm)/SiO$_2$ (3 nm). (**A**) Temporal evolutions of TR-MOKE signals, normalized by the maximum signal change, for various current densities for $t_{pw}$ = 30 ns and $\mu_0 H_x = -90$ mT. (**B**) Switching probability ($P_{SW}$) as a function of $t_{pw}$ at various current densities.

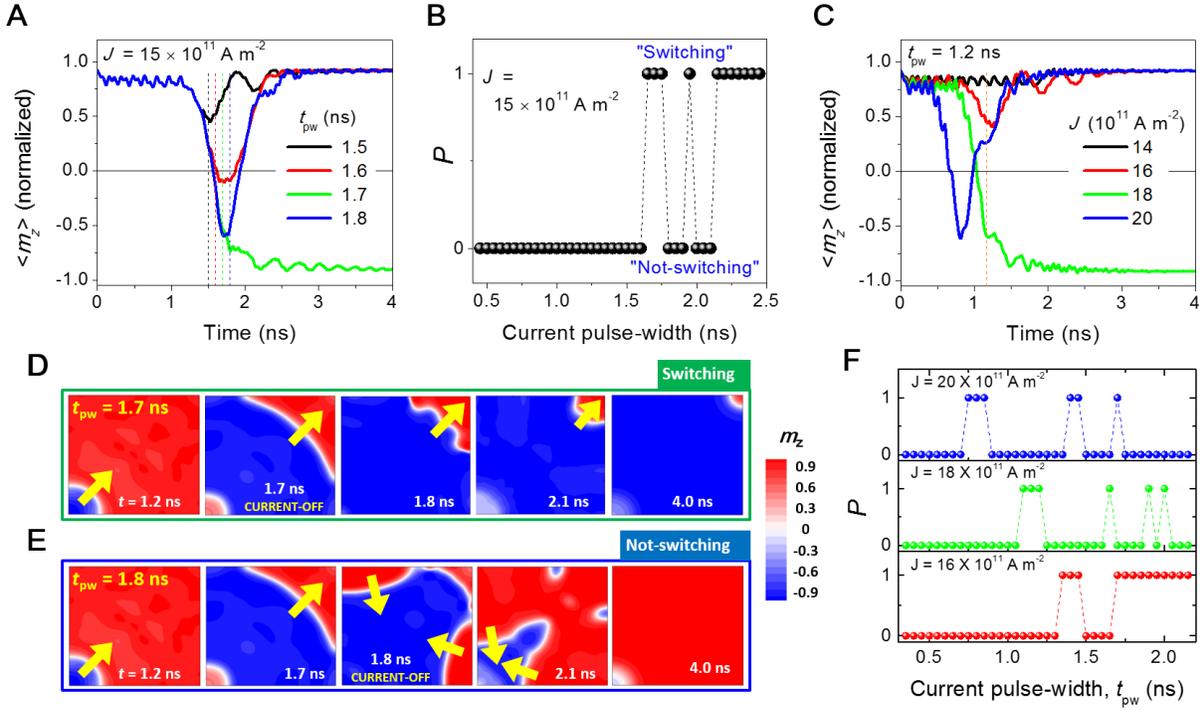

**Fig. 4. Micromagnetic simulation results for SOT switching at 0 K.** (**A**) Temporal evolutions of average $M_z/M_s$ ($<m_z>$) at various current pulse-widths. (**B**) Switching parameter ($P$, "0" = No-switching, "1" = Switching) as a function of $t_{pw}$ for $J = 15 \times 10^{11}$ A m$^{-2}$ and $\mu_0 H_x = -200$ mT. (**C**) Temporal evolutions of $<m_z>$ at various current densities for $t_{pw} = 1.2$ ns and $\mu_0 H_x = -200$ mT. Snapshots of magnetization configuration ($m_z = M_z/M_s$) at time $t$ for $t_{pw} = 1.7$ ns (**D**) and 1.8 ns (**E**). Yellow arrows show the direction of domain wall motion. (**F**) Switching parameter $P$ as a function of $t_{pw}$ for various current densities ($J = 16, 18, 20 \times 10^{11}$ A m$^{-2}$).

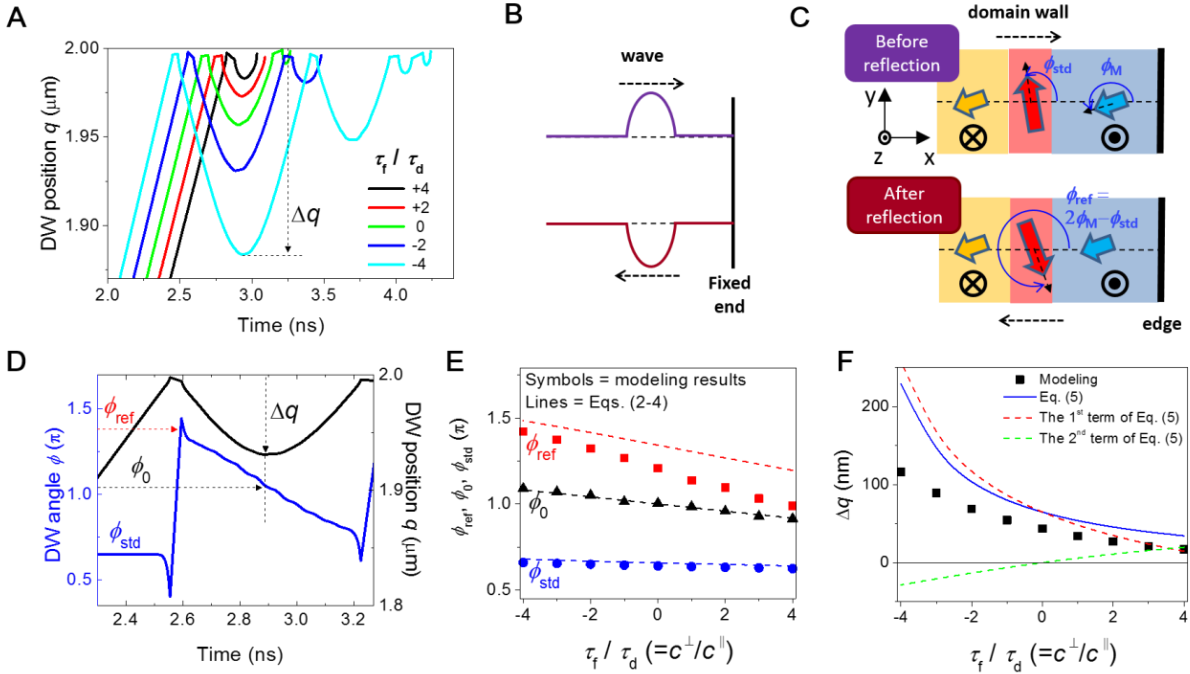

**Fig. 5. Domain wall reflection in one-dimensional model.** (**A**) Temporal evolutions of the domain wall position $q$ at various ratios of field-like torque to damping-like torque ($\tau_f/\tau_d = c^\perp/c^\parallel$) for $c^\parallel = -0.07$, $J = 6 \times 10^{11}$ A m$^{-2}$ and $\mu_0 H_x = -200$ mT. $\Delta q$ is the distance for the backward motion of a reflected domain wall. Schematic illustrations of the reflection of a transverse wall at a fixed end (**B**), and the reflection of a domain wall at an edge (**C**). (**D**) Temporal evolutions of domain wall angle $\phi$ and domain wall position $q$ for $\tau_f/\tau_d = -2$. Domain wall angles $\phi_{std}$, $\phi_M$, $\phi_{ref}$, and $\phi_0$ are defined in (**C**) and (**D**) (see the text for details). (**E**) $\phi_{ref}$, $\phi_0$, and $\phi_{std}$ as a function of $\tau_f/\tau_d$. (**F**) $\Delta q$ as a function of $\tau_f/\tau_d$.

22